\documentclass[onecolumn, 11 pt]{article}

\usepackage{graphicx}      
\usepackage{pict2e}
\usepackage{epstopdf}
\usepackage{natbib}        

\usepackage{amsmath, amsfonts, amssymb}
\usepackage{color}
\usepackage{algorithm,tikz}

\usepackage{geometry}
\newcommand{\W}{\mathcal{W}}
\newcommand{\V}{\mathcal{V}}
\newcommand{\R}{\mathbb{R}}
\newcommand{\Y}{\mathcal{Y}}
\newcommand{\T}{^{\mbox{\tiny \sf T}}}
\newcommand{\tr}{\mathrm{tr}}
\newcommand{\q}{Q_t(\sigma)}
\newcommand{\qq}{Q_{t|t-1}(\sigma)}
\newcommand{\qed}{\blacksquare}

\newcommand{\rev}[1]{ #1 }
\newcommand{\revv}[1]{ #1 }
\newtheorem{problem}{Problem}
\newtheorem{remark}{Remark}
\newtheorem{lem}{Lemma}
\newtheorem{thm}{Theorem}
\newtheorem{prop}{Proposition}
\newenvironment{pf}{%
    \textit{Proof:}%
}{%
    \\
}

\title{\vspace{-1 in} \LARGE \bf Sensor Scheduling for Linear Systems:\\ A Covariance Tracking  Approach \footnote{
Email Addresses: \texttt{dmaity@uncc.edu, dhartma2@umd.edu, baras@umd.edu}.
}} 

\author{Dipankar Maity$^*$, David Hartman$^\dagger$, and John S. Baras$^\dagger$}

\date{\tiny $*$ Department of Electrical and Computer Engineering, University of North Carolina at Charlotte, NC, 28223, USA\\
$\dagger$ Department of Electrical and Computer Engineering, University of Maryland at College Park, MD, 20742, USA} 

\begin{document}

\maketitle


\begin{abstract}                
We consider the classical sensor scheduling problem for linear systems\rev{where only one sensor is activated at each time.}We show that the sensor scheduling problem has a close relation to the  \textit{sensor design} problem and the solution of a sensor schedule problem can be extracted from an equivalent sensor design problem. We propose a convex relaxation to the sensor design problem and 
a reference covariance trajectory is obtained from solving the relaxed sensor design problem.
Afterwards, a \textit{covariance tracking} algorithm is designed to obtain an approximate solution to the sensor scheduling problem using the reference covariance trajectory obtained from the sensor design problem.
While the sensor scheduling problem is NP-hard, the proposed framework circumvents this computational complexity by decomposing this problem into a convex sensor design problem and a covariance tracking problem.
We provide theoretical justification and a sub-optimality bound for the proposed method using dynamic programming.
The proposed method is validated over several experiments portraying the efficacy of the framework.
\end{abstract}




\section{Introduction}
\subsection{Motivation and Prior Work}
    Advancements in network control systems,  distributed systems, and the development of multi-agent autonomous systems for surveillance require the development of efficient algorithms allocating resources to manage the sensory data originating from a large number of sensors observing different parts of a single or distributed system, see for example, \cite{evans2005networked,gupta2006stochastic,zhang2006communication,williams2007information}.
    These  problems  have  a long  history starting with  \cite{c1,athans1972determination},  and they include  static sensor scheduling problems as well as trajectory optimization scenarios for mobile sensors, e.g., \cite{williams2007information}.
    In the sensor scheduling problem, we aim to minimize an error criteria (e.g., the mean square error) where the error is dependent on sensor measurements over a fixed time horizon. We are constrained by the number of sensors that can be activated at each time. 
        This problem has many applications including estimation of spatial phenomenon in \cite{Nowak2004}, target tracking in \cite{Masazade2012}, robot navigation in \cite{Vitus2011}.
        Different methods have been applied to solve these problems. 
        For example, in \cite{Nowak2004} the scheduler is found by employing hierarchical sensor networks to trade-off the mean square error and communication cost.
        Whereas, a sparsity promoting penalty function is added to the objective function to help find a scheduler in  
         \cite{Masazade2012}, and, in  \cite{Vitus2011}, a scheduler is found by solving a novel incremental optimization problem.  

    The optimization problems solving for optimal sensor schedules are generally mixed-integer nonlinear programs, and thus, quickly become intractable. 
    Often these optimization problems do not posses any amenable structure that could be exploited to reduce their computational complexities. 
    Owing to this difficulty, a whole array of approximation  attempts have been proposed to solve these problems.
     The work of   \cite{c1} proposed a solution that checks all possible sensor schedules, whereas,  
     \cite{c2} devised a solution that prunes the exponentially sized search tree to reduce the search space at the expense of added computation due to pruning. 
     \cite{c3} relaxed the problem into a convex optimization problem, a heuristic that often works well in practice. 
     \cite{c4} modeled the sensor scheduling problem as a partially observable Markov decision process (POMDP) and proposed approximate solutions to solve this POMDP. 
     \rev{A stochastic optimization based solution for an infinite-horizon steady state problem is addressed by \cite{gupta2006stochastic}.}
    Greedy solutions to the sensor scheduling problem have been proposed as well, such as \cite{c5}. 
     \cite{c6}  propose a greedy solution that also includes integer programming.
    Some of these greedy solutions leverage ideas from submodularity, e.g., \cite{c8}.
    These approaches also include optimizing a slightly different objective function (e.g., convexification of the objective). 
    Additionally, \cite{c11} has proposed an optimal sensor schedule by restricting the scheduling in the class of periodic functions. 
    The existence of periodic sensor scheduling has been proven in \cite{orihuela2014periodicity} where a sub-optimal one-step ahead strategy is thoroughly studied as a possible example of observation scheme.
    
    \subsection{Contribution of This Work}
We revisit the sensor scheduling problem of linear Gaussian systems and recast it as a sensor design problem, which, to the best of our knowledge, has not been explored in the past research. 
\rev{ While in a sensor scheduling problem we search for the optimal schedule for a given set of sensors, in a sensor design problem (details are provided in Section~3.1) we design the optimal sensors. 
These two classes of problems are treated differently as their primary objectives are different. 
However, sensor scheduling problems have a close connection to sensor design problems and we show that the former can be expressed as a special case of the latter.}We further demonstrate that a class of sensor design problems can be solved by convex optimization.\rev{Using this derived connection between sensor scheduling problems and sensor design problems, an approximate solution of a scheduling problem to select one sensor at any time is constructed.}While some of the prior works e.g., \cite{c3} and \cite{c8} modify the cost function of a scheduling problem to make it convex, we show that the sensor design problem is already a convex optimization problem when the sensor parameters are restricted to a convex set. 
    In this approach, we thus avoid the need to solve mixed integer programming problems which are inherent to sensor scheduling problems.
    
    The contributions of this work are as follows: First, we show that there is a one-to-one connection between sensor scheduling problems and  sensor designing problems. 
    Second, we show that sensor designing problems are indeed  convex when the optimization parameters lie within convex sets. 
    Third, from the equivalent design problem of a scheduling problem, we obtain a \textit{reference covariance trajectory} that is used by our tracking algorithm (Algorithm~\ref{A:algo}) to find a sensor schedule.\rev{
    Fourth, we use an approximate dynamic programming based argument to provide guarantees and a sub-optimality bound of our algorithm.
   Finally, through numerical experiments, we demonstrate the efficacy of our proposed method over some existing methods, e.g., \cite{gupta2006stochastic} and \cite{c8}.}\\

    The rest of the paper is structured as follows: In Section~\ref{S:formulation}, the problem is formulated as a nonlinear integer program.
    In Section~\ref{S:design}, we propose a solution to this problem where we construct a sensor design problem, and then in Section~\ref{S:scheduling}, we provide a covariance-tracking algorithm to obtain the solution of the original sensor scheduling problem from the solution of the sensor design problem.
    We analyze the performance of our algorithm in Section~\ref{S:DP} using dynamic programming based arguments.
    Numerical analysis on the performance  of our approach is provided in Section~\ref{S:simulation}.
    Finally, we conclude this article in Section~\ref{S:conclusion}.

\section{Problem Formulation}\label{S:formulation}
We consider a linear system of the form
\begin{align}\label{E:dyn}
    X_{t+1}=A_t X_t+W_t,
\end{align}
where $X_t\in \R^n$,  $X_0 \sim \mathcal{N}(\mu_0,\Sigma_0)$ is the initial state, $\{W_t\}_{t=0}^T$ is an i.i.d sequence of Gaussian random variable with $W_0 \sim \mathcal{N}(0,\W)$, and $W_t$ is independent of $X_0$ (denoted as $W_t\!\perp\!\!\!\perp \!X_0$)  for all $t$.
 The dynamical system \eqref{E:dyn} is equipped with $N $ sensors which are described by
\begin{align} \label{E:measurement}
    Y^i_t=C^i_t X_t+V_t^i, \quad i\in \mathsf{N}\triangleq\{1,\ldots, N\},
\end{align}
where $\{V^i_t\}_{t=0}^T$ is an i.i.d sequence of Gaussian random variables with $V^i_0\sim \mathcal{N}(0,\V^i)$. 
Furthermore, for all $t,s$ and $i\ne j \in \mathsf{N}$,  $V^i_t$ and $V^j_s$ are independent, i.e., $V^i_t \perp\!\!\!\perp V^j_s$, and also  $V^i_t \perp\!\!\!\perp X_0$, $V^i_t \perp\!\!\!\perp W_s$.

Only one out of the $N$ sensors are used at any time to obtain the measurements, which are then used in estimating the state of the system \eqref{E:dyn}.
Let $\sigma:[0,T]\to \mathsf{N}$ be a sensor schedule function such that $\sigma(t)=i$ denotes that the $i$-th sensor is used at time $t$ to obtain the measurement $Y^i_t$ corresponding to sensor $i$.
Thus, the received measurements up until time $t$ can be represented as $\Y_t(\sigma)\triangleq\{Y^{\sigma(0)}_0,\ldots,Y^{\sigma(t)}_t\}$.
For a given schedule $\sigma$, the estimation error and the estimation error covariance  at time $t$ are defined as $e_t(\sigma) \triangleq X_t-\mathbb E[X_t|\Y_t(\sigma)] $ and $P_t(\sigma) \triangleq \mathbb{E}[e_t(\sigma)e_t(\sigma)\T]$, respectively.
The objective is to find a sensor schedule $\sigma$ to minimize a cumulative expected quadratic error $\sum_{t=0}^T\mathbb E[e_t(\sigma)\T e_t(\sigma)]$ over a finite horizon $[0,T]$, that is, to minimize $ \sum_{t=0}^{T}\tr(P_t(\sigma))$.

We define the following two matrix valued functions to maintain brevity in the subsequent analysis.
\begin{subequations} \label{E:gh}
\begin{align}
   & g_t(i,M)=M -M{C_t^{i}}\T (C_t^i M {C_t^i}\T +\V^i)^{-1}{C_t^i}M, \label{E:g}\\
    &h_t(M)=A_{t-1} M A_{t-1}\T+\W. \label{E:h}
\end{align}
\end{subequations}
From Kalman filtering theory, we obtain that, for all $t$,
\begin{subequations}
\begin{align}
    &P_t(\sigma)=g_t(\sigma(t),P_{t|t-1}(\sigma)), \label{E:pt}\\
   & P_{t|t-1}(\sigma)=h_t(P_{t-1}(\sigma)),~~~~P_{0|-1}=\Sigma_0.\label{E:p1t}
\end{align}
\end{subequations}
The optimal sensor scheduling problem is as follows.
\begin{problem}[Sensor scheduling problem] \label{P:prob1}
Given a system \eqref{E:dyn}-\eqref{E:measurement}, find a  schedule $\sigma:[0,T]\to \mathsf{N}$ that solves the following optimization problem:
\begin{align*}
    \min &\sum_{t=0}^{T}\tr(P_t(\sigma))\\
    \text{subject to }& P_t(\sigma)=g_t(\sigma(t),P_{t|t-1}(\sigma)), \nonumber\\
    &P_{t|t-1}(\sigma)=h_t(P_{t-1}(\sigma)),\quad P_{0|-1}=\Sigma_0,
\end{align*}
\rev{with variables $\sigma, P_t, P_{t\mid t-1}.$}
\end{problem}
Problem~\ref{P:prob1} is combinatorial in nature due to the discrete mapping of the scheduling function $\sigma(\cdot)$, and generally, it is NP-hard, see e.g., \cite{c8}. 
Majority of the prior works rely on integer programming or relaxations  to solve Problem~\ref{P:prob1}.
In this work, we propose an efficient sub-optimal solution that studies the problem from the perspective of sensor-design rather than sensor scheduling.


\section{Optimal Sensor Schedule} \label{S:solution}
In order to solve Problem \ref{P:prob1},  we will construct a computationally inexpensive simplified problem that provides a sub-optimal solution to Problem \ref{P:prob1}.
In this section, instead of looking for a schedule $\sigma(\cdot)$, we focus on a seemingly different problem that seeks to design sensors to be used over the horizon $[0,T]$.

Before starting our discussion on the sensor design problem and its relation with the scheduling problem of Problem~\ref{P:prob1}, let us define two matrices $Q_t$ and $Q_{t|t-1}$ (\rev{information matrices for Kalman filtering}) as follows
\begin{align*}
Q_t(\sigma)\triangleq P_t^{-1}(\sigma),\quad Q_{t|t-1}(\sigma) \triangleq P^{-1}_{t|t-1}(\sigma).
\end{align*}
Therefore, from \eqref{E:g} and \eqref{E:pt}, we may write:
\begin{align*}
    Q_t^{-1}(\sigma)=~&Q_{t|t-1}^{-1}(\sigma)-Q_{t|t-1}^{-1}(\sigma)C_t\T({\sigma})S^{-1}_t(\sigma)C_t(\sigma)Q_{t|t-1}^{-1}(\sigma),
\end{align*}
where $S_t(\sigma) \triangleq C_t(\sigma)P_{t|t-1}(\sigma)C_t(\sigma)\T +\V_t(\sigma)$ and $C_t(\sigma) \triangleq C^{\sigma(t)}_t$, $\V_t(\sigma) \triangleq  \V^{\sigma(t)}$.
Using Woodbury equality\footnote{$(A+UCV)^{-1}\triangleq A^{-1}-A^{-1}U(C^{-1}+VA^{-1}U)^{-1}VA^{-1}$}\!, one may equivalently write
\begin{align}
    \q = \qq +C_t\T(\sigma)\V_t^{-1}(\sigma) C_t(\sigma).
\end{align}

Using the new variables $\q$ and $\qq$, Problem~\ref{P:prob1} is rewritten as Problem~\ref{P:prob2}.
\begin{problem} \label{P:prob2}
Given a system \eqref{E:dyn}-\eqref{E:measurement}, find a  schedule $\sigma:[0,T]\to \mathsf{N}$ that solves the following problem:
\begin{align*}
    \min &\sum_{t=0}^{T}\tr(P_t(\sigma)) \nonumber\\
    \text{subject to }& P_t(\sigma)=Q_t^{-1}(\sigma), \quad P_{t|t-1}(\sigma)=h(P_{t-1}(\sigma)), \nonumber\\
    &\q = \qq +C_t\T(\sigma)\V_t^{-1}(\sigma) C_t(\sigma),\nonumber \\
     &Q_{t|t-1}(\sigma)=P_{t|t-1}^{-1}(\sigma), \quad Q_{0|-1}=\Sigma_0^{-1},\nonumber
\end{align*}
\rev{with variables $\sigma, P_t, P_{t\mid t-1}, Q_t,  Q_{t\mid t-1}.$}
\end{problem}
Although, the formulations in Problem~\ref{P:prob1} and Problem~\ref{P:prob2} may appear different as their constraints are different, one can verify that these two problems are equivalent, and thus, by solving one, we can recover the solution for the other.
Next, we show the advantage of using Problem~\ref{P:prob2} over Problem~\ref{P:prob1} for solving  the sensor scheduling problem by discussing a sensor design problem.

\subsection{Sensor Design Problem} \label{S:design}
In this section, we focus on solving a sensor design problem which is closely related to the sensor scheduling problem and provides a reasonable heuristic for solving the latter problem.
To that end, a related sensor design problem is presented in Problem~\ref{P:design2}.
\begin{problem} \label{P:design2}
Given a system \eqref{E:dyn}, and the sets $\mathsf{C}_t$ and $\mathsf V_t$, design a linear sensor $Y_t=C_tX_t+V_t$ where $V_t\sim\mathcal{N}(0,\V_t)$,  to solve the following problem:
\begin{align}
    \min&\sum_{t=0}^{T}\tr(P_t)\nonumber \\
    \text{subject to  } &P_t=Q_t^{-1},\quad Q_t=Q_{t|t-1}+R_t, \nonumber\\
    &Q_{t|t-1}=\left(h_t(P_{t-1})\right)^{-1},\quad Q_{0|-1}=\Sigma_0^{-1}, \nonumber\\
    &R_t=C_t \T \V_t^{-1}C_t,~~~ C_t \in \mathsf{C}_t, ~~\V_t \in \mathsf{V}_t, \nonumber
\end{align}
\rev{with variables $ P_t,  Q_t,  Q_{t\mid t-1}, R_t, C_t, \V_t$.}
\end{problem}
\begin{remark} \label{R:design_schedule}
By replacing the constraints $C_t\in \mathsf{C}_t, \V_t \in \mathsf{V}_t$ with $(C_t,\V_t) \in\{(C^i_t,\V^i)\}_{i\in \mathsf{N}} $ in Problem \ref{P:design2}, we recover Problem~\ref{P:prob2}.
\end{remark}
We may further relax the constraints in Problem~\ref{P:design2} to their equivalent linear matrix inequalities (LMIs)  to obtain a relaxed problem.
By using $P_t \succeq Q^{-1}_t$ and $Q_{t|t-1} \preceq \left(h_t(P_{t-1})\right)^{-1}$, and after some simplifications via using Schur Complement, we obtain the following relaxed problem.
%
\begin{problem} \label{P:design3}
For the given dynamical system \eqref{E:dyn}, design a sensor of the form $Y_t=C_tX_t+V_t$, where $V_t\sim\mathcal{N}(0,\V_t)$, to solve the following problem:
\begin{align*}
    \min&\sum_{t=0}^{T}\tr(P_t)\nonumber \\
    \text{subject to  } &Q_t=Q_{t|t-1}+R_t, \nonumber\\
    &\begin{bmatrix} P_t & I \\ I & Q_t \end{bmatrix}  \succeq  0,\\
    &\begin{bmatrix}\W^{-1}-Q_{t|t-1} & \W^{-1}A_t\\ A\T_t \W^{-1} &  Q_{t-1}+A\T_t \W^{-1}A_t \end{bmatrix} \succeq 0, \nonumber\\
    & R_t \in \mathsf{R}_t, \quad Q_{0|-1}=\Sigma_0^{-1} \nonumber,
\end{align*}
with variables $ P_t, Q_t,  Q_{t\mid t-1}, R_t$, where
 $\mathsf{R}_t=\{C_t \T \V_t^{-1}C_t \mid C_t \in \mathsf{C}_t, ~~\V_t \in \mathsf{V}_t\}$ is a given set of positive semidefinite matrices which is constructed from the sets $\mathsf{C}_t$ and $\mathsf V_t$.
\end{problem}
Note that the constraint $R_t \in \mathsf{R}_t$ in Problem~\ref{P:design3} is an equivalent representation of the constraints $R_t=C_t \T \V_t^{-1}C_t,~~~ C_t \in \mathsf{C}_t, ~~\V_t \in \mathsf{V}_t$ in Problem~\ref{P:design2}.
Once $\{R_t\}_{t=0}^T$ is found by solving Problem~\ref{P:design3}, $\{C_t,\V_t\}_{t=0}^T$ can be found by solving 
\begin{align*}
    C_t\T \V_t^{-1}C_t=  R_t,
\end{align*}
for all $t=0,\ldots,T.$ 
While Problem~\ref{P:design3} is a relaxation of Problem~\ref{P:design2}, the following theorem states that an optimal solution of Problem~\ref{P:design3} is also an optimal solution for Problem~\ref{P:design2}.
\begin{thm} \label{pr:equivalnece}
An optimal solution of the relaxed problem (Problem \ref{P:design3}) is also an optimal solution of the original problem (Problem \ref{P:design2}), and vice-versa.
\end{thm}
\begin{pf}
Firstly, due to the relaxations, any feasible solution of Problem~\ref{P:design2} is a feasible solution for Problem~\ref{P:design3}, and hence the optimal solution of Problem~\ref{P:design2} is a feasible solution for Problem~\ref{P:design3}.
The theorem is proved once we show that for every feasible solution of Problem \ref{P:design3} there exists a feasible solution for Problem \ref{P:design2} that produces the same, if not a lesser, objective value.
In order to prove this, let us consider $\{P_t,Q_t,Q_{t|t-1}\}$ to denote a feasible solution of Problem \ref{P:design3} 
and let us construct a new tuple $\{\bar P_t,\bar Q_t,\bar Q_{t|t-1}\}$, for all $t$, as follows: 
\begin{align}\label{E:solution_bar}
\begin{split}
    &\bar{Q}_{0|-1}=Q_{0|-1}\quad  \bar R_t=Q_t-Q_{t|t-1} ,\quad \bar{P}_t=\bar{Q}_t^{-1} , \\
    &\bar{Q}_t=\bar Q_{t|t-1}+\bar R_t, \quad \bar{Q}_{t+1|t}=\left(h_{t+1}(\bar P_{t})\right)^{-1}.
    \end{split}
\end{align}
Based on this construction of $\{\bar P_t,\bar Q_t,\bar Q_{t|t-1}\}$, one can verify using mathematical induction that $\bar P_t\preceq P_t$, $\bar{Q}_t \succeq Q_t$ and $\bar Q_{t|t-1} \succeq Q_{t|t-1}$ for all $t$.
Matrix $\bar R_t$, as defined in \eqref{E:solution_bar}, satisfies $\bar R_t\in \mathsf R_t$.
Therefore, based on \eqref{E:solution_bar}, one can conclude that $\{\bar P_t,\bar Q_t,\bar Q_{t|t-1}\}$ is a feasible solution of Problem~\ref{P:design2} since they satisfy all the constraints of Problem~\ref{P:design2}. 
Furthermore, since $\bar P_t\preceq P_t$, it then follows that $\sum_{t=0}^T\tr(\bar P_t) \le \sum_{t=0}^T\tr(P_t)$.
Thus, an optimal solution of Problem~\ref{P:design3} is a feasible solution of Problem~\ref{P:design2}, and vice-versa.
This completes the proof. \hfill $\qed$
\end{pf}

Due to Theorem \ref{pr:equivalnece}, the LMI-based relaxations introduced in Problem~\ref{P:design3} do not affect the optimality since an optimal solution of the relaxed problem is also optimal for the original problem.
This is a key advantage of this approach over existing methods. 
It is noteworthy that the LMI-based relaxations retain the optimality of a sensor-design problem.
\revv{Furthermore, Problem~\ref{P:design3} is a mixed integer semidefinite program and one could use efficient numerical techniques, e.g., \cite{gally2018framework} to solve it directly. }

Note that Problem~\ref{P:design3} is convex when $\mathsf{R}_t$ is a convex set for all $t$. 
Moreover, if $\mathsf{R}_t$ is a convex hull of a set of $\ell$ matrices $\{R^1_t,\ldots,R^\ell_t\}$ for all $t$, then we can replace the constraint $R_t \in \mathsf{R}_t$ with the constraints $R_t=\sum_{i=1}^\ell \theta^i_t R^i_t$,~~ $\theta^i_t\in [0,1]$ and $\sum_{i=1}^\ell \theta^i_t=1$.
In this case, Problem~\ref{P:design3} can be further simplified to Problem~\ref{P:design_SDP}.
\begin{problem}\label{P:design_SDP}
For the given dynamical system \eqref{E:dyn}, design a sensor of the form $Y_t=C_tX_t+V_t$, where $V_t\sim\mathcal{N}(0,\V_t)$, to solve the following problem:
\begin{align}
    \min&\sum_{t=0}^{T}\tr(P_t)\nonumber \\
    \text{subject to  }\qquad &Q_t=Q_{t|t-1}+\sum_{i=1}^\ell \theta^i_t R^i_t,\quad Q_{0|-1}=\Sigma_0^{-1} \nonumber\\
    &\begin{bmatrix} P_t & I \\ I & Q_t \end{bmatrix} \succeq 0,\nonumber\\ 
    & \begin{bmatrix}\W^{-1}-Q_{t|t-1} & \W^{-1}A_t\\ A\T_t W^{-1} &  Q_{t-1}+A\T_t \W^{-1}A_t \end{bmatrix} \succeq 0, \nonumber\\
   &  \sum_{i=1}^\ell \theta^i_t =1, \quad 0 \le \theta^i_t \le 1 \nonumber.
\end{align}
\rev{with variables $\theta_t, P_t,  Q_t,  Q_{t\mid t-1}$.}
\end{problem}
At this point, we are ready to address the sensor scheduling problem and its connection to Problems~\ref{P:design2}-\ref{P:design_SDP}.

\subsection{Sensor Scheduling Problem}\label{S:scheduling}
The sensor scheduling problem can be viewed as a sensor design problem if we restrict the design variables $(C_t,\V_t)$ to be one of the elements in the set $\{({C}^i_t,\V^i)\}_{i\in \mathsf{N}}$ as mentioned in Remark \ref{R:design_schedule}. 
Equivalently, if we restrict $\mathsf R_t = \{R^i_t\}_{i \in \mathsf{N}}$  in Problem~\ref{P:design3}, where $R^i_t={C^i_t}\T{\V^i}^{-1} C^i_t$, then we recover a solution to Problem~\ref{P:prob1}.
However, solving Problem~\ref{P:design3} with the non-convex constraint $R_t\in \{R^i_t\}_{i \in \mathsf{N}}$ is computationally expensive despite the availability of efficient numerical techniques, e.g., \cite{gally2018framework}. 
Therefore, we relax the constraint $R_t\in \{R^i_t\}_{i \in \mathsf{N}}$ as $R_t\in co\left( \{R^i_t\}_{i \in \mathsf{N}}\right)$ where $co(\cdot)$ denotes the convex hull operation.
With this convex hull relaxation approach, the relaxed sensor scheduling problem becomes exactly the same as Problem~\ref{P:design_SDP}.

By solving a relaxation of Problem~\ref{P:prob1}, as presented in Problem~\ref{P:design_SDP}, one obtains the variables $\{\{{\theta^i_t}^o\}_{i\in \mathsf{N}}\}_{t=0}^T$, or equivalently $R^o_t=\sum_{i=1}^N{\theta^i_t}^o R^i_t$ and the associated $P^o_t, Q^o_t$ and $Q^o_{t|t-1}$.
If, ${\theta^i_t}^o\in \{0,1\}$ for all $i$ and $t$, then this relaxed optimal solution $\{\{{\theta^i_t}^o\}_{i\in \mathsf{N}}\}_{t=0}^T$ is an optimal schedule for Problem~\ref{P:prob1}.
However, in general, the obtained ${\theta^i_t}^o$ are not binary-valued, and hence the solution of Problem \ref{P:design_SDP} may not readily be useful as a solution to Problem~\ref{P:prob1}. 
Therefore, we propose a tracking-based algorithm to use the solution of Problem~\ref{P:design_SDP} as a guide to construct a sub-optimal solution to Problem~\ref{P:prob1}.

\begin{algorithm}[h]
\textbf{Input} $\leftarrow$ $\{P^o_t\}_{t=0}^T$, $P_{0|-1}=\Sigma_0$,\\
\textbf{for} $t=0:T$\\
$~~~~~~ M_t(i)\leftarrow g_t(i,P_{t|t-1}),\quad i\in \mathsf{N}$,\\
$~~~~~~\sigma(t)\leftarrow \arg\!\min_i \|P^o_t-M_t(i)\|_F$\\
$~~~~~~P_t\leftarrow g_t(\sigma(t),P_{t|t-1}),$\\
$~~~~~~P_{t+1|t}\leftarrow h_t( P_t),$\\
\textbf{end}\\
\textbf{Output} $\leftarrow \sigma$.
\caption{Covariance Tracking Algorithm} \label{A:algo}
\end{algorithm}
Algorithm~\ref{A:algo} takes the solution $\{P^o_t\}_{t=0}^T$ obtained from solving Problem~\ref{P:design_SDP} as an initial guess, and initiates $P_{0|-1}$ at $\Sigma_0$ as required by \eqref{E:p1t}. The notation $\|\cdot\|_F $ in Algorithm~\ref{A:algo} represents the Frobenius norm.
The algorithm produces a covariance trajectory $\{P_t\}_{t=0}^T$ that is \textit{close} to the reference trajectory $\{P^o_t\}_{t=0}^T$ in Frobenius norm.

The reasoning behind the construction of Algorithm~\ref{A:algo} is to keep the error covariance $P_t(\sigma)$ close to $P_t^o$, since $P_t^o$ is the best covariance one could possibly obtain given the set of sensors.
The algorithm is reminiscent of a \textit{trajectory-tracking} problem where $P^o_t$ serves as the reference trajectory. 
In the following we formally provide technical justifications of using such a heuristic method and its merits using dynamic programming based arguments.

\subsection{Dynamic Programming and Optimality Guarantees}\label{S:DP}
Let us denote the value function associated with Problem~\ref{P:prob1} to be $U_t$, which is given as follows
\begin{align} \label{E:value}
    U_t(P)=\min_{ P_{t|t-1}=P,~\{\sigma(k)\}_{k=t}^{T}}\sum_{k=t}^{T}\tr(P_k(\sigma)),
\end{align}
where $P$ is a positive semidefinite matrix.
Similarly, we denote the value function associated with the SDP relaxation of Problem~\ref{P:prob1} (equivalent to Problem \ref{P:design_SDP} with $\ell=N$ and $R^i_t={C^i_t}\T {\V^{i}}^{-1}C^i_t$) by $U^o_t$: 
\begin{align}
    U_t^o(P)=\min_{P_{t|t-1}\!=P,~\{\{\theta^i_k\}_{i\in \mathsf{N}}\}_{k=t}^{T} }\sum_{k=t}^{T}\tr(P_k(\sigma)).
\end{align}
The difference between $U_t$ and $U^o_t$ is that the feasible choice of a sensor at time $t$ for $U_t$ has to be one of the $\{R^i_t\}_{i\in \mathsf{N}}$ (or equivalently $\{C^i_t,\V^i_t\}_{i\in \mathsf{N}}$), whereas the feasible choice of a sensor for $U^o_t$ is any of the sensors that lie within the convex hull of $\{R^i_t\}_{i\in \mathsf{N}}$.
Therefore,  $U^o_t(P) \le U_t(P)$ for all symmetric $P\succeq 0$. 
In what follows, we will suppress the $P_{t|t-1}=P$ constraint in the definitions of the value function to maintain notational brevity.

From dynamic programming, one may write
\begin{align*}
    U_t(P)=&\min_{\sigma(t)}\left(\tr(P_t(\sigma))+U_{t+1}(P_{t+1|t}(\sigma))\right)\\
    =&\min_{\sigma(t)}\left(\tr\big(g_t(\sigma(t),P)\big)+U_{t+1}\Big( h_{t+1}\big(g_t(\sigma(t),P)\big)\Big)\right),\\
    U_T(P)=&\min_{\sigma} \tr\big(g_T(\sigma,P)\big).
\end{align*}
In the following we will exploit some of the properties of  $U_t$ and the solutions obtained from solving the SDP relaxation ($P^o_t, Q^o_t$ and $P^o_{t|t-1}$) to solve for an approximate value function associated with \eqref{E:value}.

Before proceeding, let us present some useful properties of the map $g_t(\cdot,\cdot)$  defined in \eqref{E:gh} which will assist us in our subsequent analyses. 
With a slight variation to Lemma~1-e from \cite{sinopoli2004kalman}, one can prove that, for any fixed $i\in \mathsf{N}$, $g_t(i,M)$ is concave in $M$.
Furthermore, we can characterize the derivative of the function $g_i(i,\cdot)$ by the following lemma.
\begin{lem}[\cite{c2}] \label{L:g_concave}
For each $i\in \mathsf{N}$ and for any positive semi-definite matrices $M,L$, it follows that
\begin{align}
     \frac{dg_t(i,M+\epsilon L)}{d\epsilon}\Big|_{\epsilon=0}= H_t(i,M)L H_t(i,M)\T,
\end{align}
where $H_t(i,M)=(I-M{C^i_t}\T(C_t^i M {C_t^i}\T +\V^i)^{-1}C^i_t)$.
\end{lem}
\begin{pf}
Let us define $\tilde{g}_t(i,M)=(C_t^i M {C_t^i}\T +\V^i)^{-1}$, and therefore,
\begin{align*}
    \frac{d \tilde{g}_t(i,M+\epsilon L)}{d\epsilon}=-\tilde{g}_t(i,M+\epsilon L)C^i_t L {C_t^i}\T\tilde{g}_t(i,M+\epsilon L).
\end{align*}
Using \eqref{E:g} and after some simplifications, we obtain
\begin{align*}
     \frac{d {g}_t(i,M+\epsilon L)}{d\epsilon}\Big|_{\epsilon=0}=H_t(i,M) L H_t(i,M)\T. \hfill \qed
\end{align*} 
\end{pf}
The following proposition shows that the value function $U^o_t$ is locally Lipschitz, which is an important component in constructing Algorithm~\ref{A:algo}.
\begin{prop}\label{Pr:Lipsc}
For any two symmetric positive semidefinite matrices $P$ and $Q$ with bounded Frobenius norms, and for all $t=0,\ldots, T$, there exists a constant $K>0$, such that
\begin{align}
    \|U^o_t(P)-U^o_t(Q)\|\le K\|P-Q\|_F.
\end{align}
\end{prop}
\begin{pf}
We prove this in an inductive way. 
Let us first consider $t=T$, and hence,
\begin{align*}
    U_T^o(P)-U_T^o(Q)=&\min_{\{\theta^i\}_{i\in \mathsf{N}}}\! \tr\big(g_T(\theta,P)\big)-\!\!\min_{\{\theta^i\}_{i\in \mathsf{N}}}\!\tr\big(g_T(\theta,Q)\big)\\
    \overset{(a)}{\le} & \tr\big(g_T(\theta^*,P)-g_T(\theta^*,Q)\big)\\
    \overset{(b)}{\le}& \tr(H_T(\theta^*,Q)(P-Q)H_T\T(\theta^*,Q))\\
    \le &\|P-Q\|_F\|H_T\T(\theta^*,Q)H_T(\theta^*,Q)\|_F
\end{align*}
where $\theta^*=[\theta^{1*},\ldots,\theta^{N*}]$ in $(a)$ is a minimizer of $\tr(g_T(\theta,Q))$, and (b) follows from the concavity property of the function $g_T(\theta,\cdot)$ along with Lemma~\ref{L:g_concave}. 
From the expression of $H_T(\theta^*,Q)$ in Lemma~\ref{L:g_concave}, along with the fact that $Q$ has a bounded Frobenius norm, one can verify that there exists a finite $K>0$ such that $\|H_T\T(\theta^*,Q)H_T(\theta^*,Q)\|_F \le K$. 
Therefore, 
\begin{align*}
    U_T^o(P)-U_T^o(Q)\le K\|P-Q\|_F.
\end{align*}
The inductive hypothesis can be proven in a similar way. \hfill $\qed$
\end{pf}
The following proposition states that an upper bound on $U_t$ is found from $U_t^o$. 
\begin{prop} \label{Pr:Ualpha}
For any time $t$ and $P\succeq 0$ with bounded Frobenius norm, there exists a finite $\alpha>0$ such that
\begin{align*}
    U_t(P) \le U^o_t(P)+\alpha.
\end{align*}
\end{prop}
Based on these propositions, we are now ready to perform an approximate dynamic programming with the value function $U_t(P)$ to design a sub-optimal solution as follows.
Recall that the value function $U_t(P)$ satisfies
\begin{align*}
     U_t(P)=&\min_{\sigma(t)}\left(\tr(P_t(\sigma))+U_{t+1}(P_{t+1|t}(\sigma))\right),
\end{align*}
which can be re-written as
\begin{align*}
     U_t(P) &\le \alpha+\min_{\sigma(t)}\left(\tr(P_t(\sigma))+U^o_{t+1}(P_{t+1|t}(\sigma))\right), \\
     &\le \alpha +U^o_t(P^o_{t|t-1}) \\
     &~~+\min_{\sigma(t)}\left(\tr(P_t(\sigma))
     +U^o_{t+1}(P_{t+1|t}(\sigma))-U^o_t(P^o_{t|t-1})\right),
\end{align*}
where $P^o_{t|t-1}$ is obtained from the SDP relaxation. 
More specifically, by solving the SDP relaxation one obtains $\{P^o_t,Q^o_{t|t-1},Q^o_t\}_{t=0}^T$ and from these one can construct $P^o_{t|t-1}={Q^{o-1}_{t|t-1}}$.
Thus, we have that $U^o_t(P^o_{t|t-1})=\tr(P^o_t)+U^o_{t+1}(P^o_{t+1|t})$, and therefore,
\begin{align}  \label{eq:upper_bound}
    &U_t(P)
     \le \alpha +U^o_t(P^o_{t|t-1}) \nonumber \\ \nonumber
     &+\min_{\sigma}\left(\tr(P_t(\sigma)-P^o_t)
     +U^o_{t+1}(P_{t+1|t}(\sigma))-U^o_{t+1}(P^o_{t+1|t})\right)\\ \nonumber
     &\le \alpha +U^o_t(P^o_{t|t-1}) \\\nonumber
     &+\min_{\sigma}\left(\tr(P_t(\sigma)-P^o_t)
     +K\|P_{t+1|t}(\sigma)-P^o_{t+1|t}\|_F\right)\\ \nonumber
     &\le \alpha +U^o_t(P^o_{t|t-1}) \\ \nonumber
     &+\min_{\sigma}\left(\tr(P_t(\sigma)-P^o_t)
     +K\|A_t\|^2\|P_t(\sigma)-P^o_t\|_F\right)\\ 
     &\le \alpha +U^o_t(P^o_{t|t-1})+K_1\min_{\sigma}\|P_t(\sigma)-P^o_t\|_F,
\end{align}
where $K_1=K\|A_t\|^2+\sqrt{n}$ and we have used Proposition~\ref{Pr:Lipsc} and the fact that for any $X\in \R^{n\times n}$, $\tr(X)\le \sqrt{n}\|X\|_F$.

Thus performing the optimization $\min_{\sigma}\|P_t(\sigma)-P^o_t\|_F$ essentially minimizes an upper bound of the value function $U_t$, or equivalently, an upper bound of $\sum_{t=0}^T\tr(P_t)$.
Therefore, in essence, Algorithm~\ref{A:algo}  performs an approximate dynamic programming type optimization by minimizing an upper bound of the value function $U_t$.
The following lemma provides a sub-optimality bound of Algorithm~\ref{A:algo}. 
\begin{lem}
Let $\sigma,\sigma^*$ and $\theta^*$ denote the schedule obtained from Algorithm~\ref{A:algo}, the true optimal schedule of Problem~\ref{P:prob1}, and the solution to Problem~\ref{P:design_SDP}, respectively. Then,
\begin{align} \label{eq:optimal bound}
    \sum_{t=0}^T \tr(P_t(\sigma)) \le \sum_{t=0}^T\tr (P_t(\sigma^*))  +  \epsilon
\end{align}
where $\epsilon \triangleq \sqrt{n} \sum_{t=0}^T\frac{1- \lambda^{T+1-t}}{1 - \lambda}\beta_t + \sqrt{n}\sum_{t=0}^T \|P_t(\theta^*) - P_t(\sigma^*)\|_F$, $\beta_t \triangleq  \| g_t(\sigma^*(t),P_{t\mid t-1}(\theta^*)) - P_t(\theta^*) \|_F$ and $\lambda\triangleq \max_t \|A_{t-1}H(\sigma^*(t),P_{t|t-1}(\theta^*))\|^2$.
\end{lem}
\begin{pf} Note that, for all $t$, we have
\begin{align} \label{eq:difference}
   & \|P_t(\sigma) - P_t(\sigma^*)\|_F \le \|P_t(\sigma) - P_t(\theta^*)\|_F + \|P_t(\theta^*) - P_t(\sigma^*)\|_F, 
\end{align}
where $P_t(\sigma)$ is the estimation error covariance when the schedule $\sigma$ is used upto time $t$. 
Similar definitions for $P_t(\sigma^*)$ and $P_t(\theta^*)$ as well.
Furthermore, $P_t(\theta^*)=P_t^o$ based on the definition of $\theta^*$.
From Algorithm~\ref{A:algo}, we have 
\begin{align*}
    \|P_t(\sigma) - P_t(\theta^*)\|_F &= \min_i \|g_t(i,P_{t\mid t-1}(\sigma))- P_t(\theta^*) \|_F\\
    &\le \|g_t(\sigma^*(t),P_{t\mid t-1}(\sigma))- P_t(\theta^*) \|_F\\
    & \le \|g_t(\sigma^*(t),P_{t\mid t-1}(\sigma))- g_t(\sigma^*(t),P_{t\mid t-1}(\theta^*)) \|_F  +  \| g_t(\sigma^*(t),P_{t\mid t-1}(\theta^*)) - P_t(\theta^*) \|_F
\end{align*}
Due to the concavity of $g_t(i,\cdot)$ and from Lemma~3, we obtain 
\begin{align*}
    &\|g_t(\sigma^*(t),P_{t\mid t-1}(\sigma))- g_t(\sigma^*(t),P_{t\mid t-1}(\theta^*)) \|_F\\ &\le     \|H_t(\sigma^*(t),P_{t\mid t-1}(\theta^*))\|^2\|P_{t\mid t-1}(\sigma))-P_{t\mid t-1}(\theta^*)\|_F\\
    &\le \lambda_t  \|P_{t-1}(\sigma)  - P_{t-1}(\theta^*)\|_F
\end{align*}
where $\lambda_t \triangleq  \|A_{t-1}H(\sigma^*(t),P_{t|t-1}(\theta^*))\|^2$.
By defining $\eta_t \triangleq \|P_t(\sigma) - P_t(\theta^*)\|_F$ and $\beta_t \triangleq  \| g_t(\sigma^*(t),P_{t\mid t-1}(\theta^*)) - P_t(\theta^*) \|_F$, we obtain
\begin{align}
    \eta_t \le {\lambda}\eta_{t-1} + \beta_t, \quad \eta_0 \le \beta_0, \label{eq:eta}
\end{align}
where $\lambda=\max_t\lambda_t$.
This further gives $\eta_t \le  \sum_{k=0}^t\lambda^{t-k}\beta_k$.
Notice that $ g_t(\sigma^*(t),P_{t\mid t-1}(\theta^*))$ denotes the error covariance at time $t$ due to selecting the $\sigma^*(t)$-th sensor while the prediction covariance was $P_{t|t-1}(\theta^*)$.
Whereas, $P_t(\theta^*)$ denotes the error covariance at time $t$ starting from the same prediction covariance of $P_{t|t-1}(\theta^*)$ and using the relaxed schedule $\theta^*(t)$.
Therefore, $\beta_t= \| g_t(\sigma^*(t),P_{t\mid t-1}(\theta^*)) - P_t(\theta^*) \|_F$ denotes the covariance mismatch between the schedules $\theta^*$ and $\sigma^*$, which is caused by the integer nature of the schedule $\sigma^*$. 
Clearly, if the solution to Problem~5 is integer in nature (i.e., $\theta^*_t \in \{0,1\}$) for all $t$, then $\beta_t=0$ for all $t$. 
Also note that $A_{t-1}H(\sigma^*(t),P_{t|t-1}(\theta^*))=(A_{t-1}-K_{t-1}C^{\sigma^*(t)}_t)$, where $K_{t-1}$ is the Kalman gain associated with the schedule $\theta^*$.
When the close-loop system is stable, i.e., $(A_{t-1}-K_{t-1}C^{\sigma^*(t)}_t)$ is Hurwitz, we obtain $\lambda<1$.
From \eqref{eq:difference}, \eqref{eq:eta} and the definition of $\eta_t$, we obtain
\begin{align*}
    \sum_{t=0}^T \|P_t(\sigma) - P_t(\sigma^*)\|_F  &\le \sum_{t=0}^T\eta_t+ \sum_{t=0}^T \|P_t(\theta^*) - P_t(\sigma^*)\|_F \\
    &\le \sum_{t=0}^T\frac{\lambda^{T+1-t}-1}{\lambda-1}\beta_t + \sum_{t=0}^T \|P_t(\theta^*) - P_t(\sigma^*)\|_F,
\end{align*}
which further leads to,
\begin{align} \label{eq:optimal bound}
    \sum_{t=0}^T \tr(P_t(\sigma)) -  \sum_{t=0}^T\tr (P_t(\sigma^*)) \le  \epsilon
\end{align}
where $\epsilon \triangleq \sqrt{n} \sum_{t=0}^T\frac{\lambda^{T+1-t}-1}{\lambda-1}\beta_t + \sqrt{n}\sum_{t=0}^T \|P_t(\theta^*) - P_t(\sigma^*)\|_F$.
Equation \eqref{eq:optimal bound} provides a suboptimality bound of Algorithm~\ref{A:algo}.
From the definition of $\beta_t$, we notice that $\epsilon$ depends on the covariance mismatch between the schedules $\theta^*$ and $\sigma^*$. \hfill $\qed$
\end{pf}

From the definition of $\beta_t$, we notice that $\epsilon$ captures the covariance mismatch between the schedules $\theta^*$ and $\sigma^*$,
\revv{and consequently, portraying the effects of the relaxed sensor design problem on the overall optimality of the approach.  
If the optimal covariance $\{P_t(\sigma^*)\}_{t=0}^T$ is ``close" to the covariance from the relaxed optimization $\{P_t(\theta^*)\}_{t=0}^T$ then the suboptimality bound decreases, which is expected.
Furthermore, the bound $\epsilon$ depends on the system dimension $n$ and degrades with the system's dimension.
}


\section{Numerical Evaluations} \label{S:simulation}

We empirically evaluate the performance of our algorithm by applying it on a wide range of (randomly generated) scenarios and comparing it to (suitable modifications of) \cite{c8}  and \cite{gupta2006stochastic} and a random search algorithm.
The random search algorithm randomly generates 2000 schedules and reports the \textit{best} among these schedules.
It is noteworthy that the algorithm proposed by \cite{gupta2006stochastic} is for an infinite horizon steady-state estimation problem and here we adopt this algorithm to our finite horizon problems.

We conduct several experiments by varying the dimensions of the system. 
Four sets of experiments were conducted for seven different dimensions of  $A \in \mathbb{R}^{n\times n}$ for $n = 4,6,8,$ and $10$.
For each dimension $n$, we generate thirty scenarios by randomly generating thirty $A$ matrices with eigenvalues in the range $[1,~1.5]$. 
For each scenario, we consider four randomly chosen sensors with different numbers of dimensions (i.e., different number of rows for matrices $C_i$). 
Noise $W_t$ is chosen to have zero mean and unit variance, i.e., $ W_t \sim \mathcal N(0,I)$.
The measurement noises are $V_t^i \sim \mathcal N(0,V^i)$ where $V^i$ is a diagonal covariance matrix whose diagonal elements are chosen randomly between $0$ and $1$. 
We consider a time horizon of $T=100$.

\begin{figure}
    \centering
        \includegraphics[trim=20 0 50 20, clip, width=0.75\textwidth]{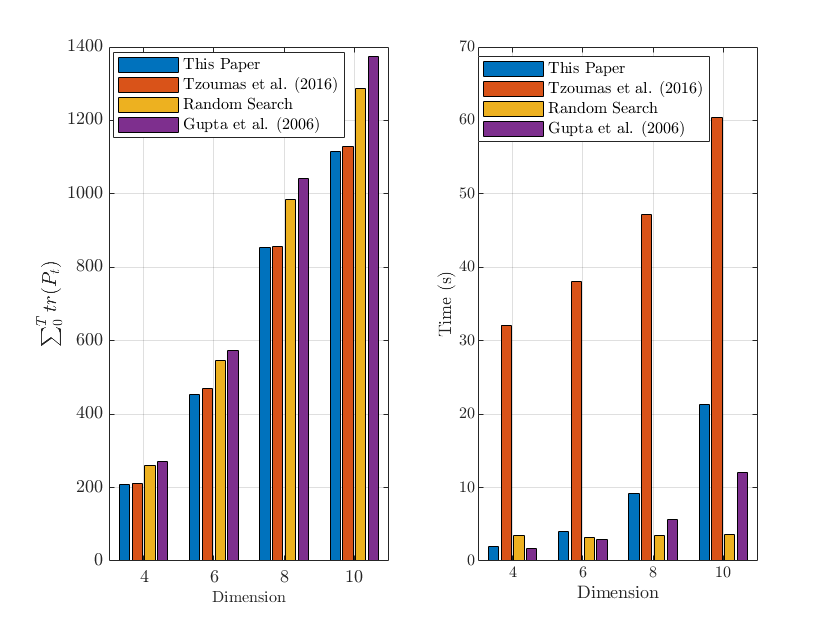}
    \caption{Plot of the cost $\sum_{t=0}^T \tr(P_t(\sigma))$ and computation times for our algorithm, \cite{c8}, \cite{gupta2006stochastic}, and a random schedule algorithm. }
 \label{F:SensorSchedule}
\end{figure}

In Figure~\ref{F:SensorSchedule} (left), we illustrate the performance of our algorithm compared to the others. 
The $x$-axis represents the dimension of the system and the $y$-axis represents the cost averaged over the randomly generated scenarios. 
As can be seen, our algorithm performs better than the other algorithms in terms of the cost $\sum_{t=0}^T\tr(P_t)$.
We also compare the run-time of these algorithms in Figure~\ref{F:SensorSchedule} (right). While the run-times for random search and \cite{gupta2006stochastic} are slightly smaller than ours, their performances are significantly worse than ours. 
On the other hand, while the performance of \cite{c8} is comparable to ours, their run-time is significantly higher compared to ours.
In this way the proposed method provides a balanced trade-off between the computation-time and the objective value.

It is worth mentioning that during our study of sensor scheduling problems, we have noticed that the optimal schedule converge to a periodic scheduling scheme  after a disproportionately shorter transient period. 
 Similar observation has also been witnessed in other works, e.g., \cite{c10} and the optimality of such periodic behaviors is proven in \cite{orihuela2014periodicity}. 
 The periodicity of the optimal policy can be further exploited in our framework to generate optimal scheduling for  infinite horizon problems and to further reduce the computational complexity for finite horizon problems. 

\subsection{Near-Optimal Performance}
\begin{figure}
    \centering
        \includegraphics[trim=20 0 0 0,clip, width=0.75 \linewidth]{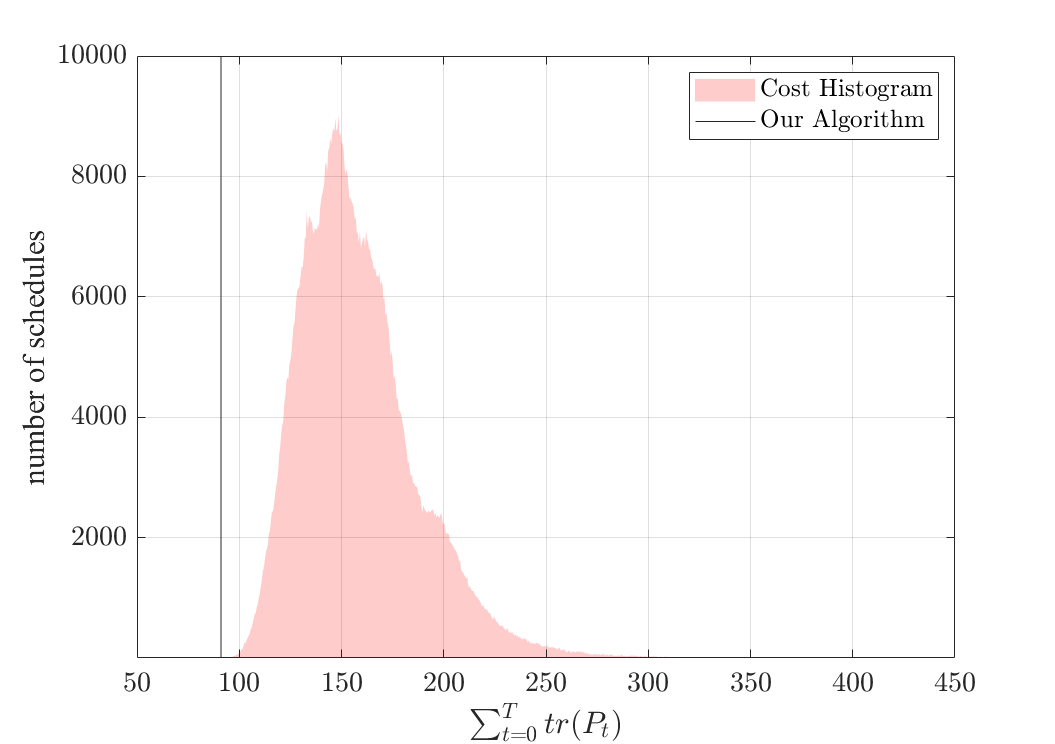}
    \caption{Plot of total costs for all permutation of schedules. 
    The gray line corresponds to the cost by using the schedule resulting from our tracking algorithm.} \label{F:ExhaustiveSearch}
\end{figure}
While in the last set of experiments we illustrated that our algorithm outperforms the other algorithms, in this section we quantitatively demonstrate how close to the true optimum our solution can be. 
To compute the optimal schedule, we need to resort to exhaustive searches, and hence, we reduce the time horizon to $T=9$ in order to maintain tractability of the exhaustive search methods.
We randomly pick one of the thirty $10$-dimensional scenarios that were used in the previous experiment. 
We have $4$ available sensors, and hence, the total number of possible schedules are $4^{10}$.
We compare the performances of all permutations of schedules ($4^{10}$ of them) to the performance resulting from our algorithm by plotting a histogram in Figure~\ref{F:ExhaustiveSearch}.
In Figure~\ref{F:ExhaustiveSearch}, the $x$-axis represents the cost ($\sum_{t=0}^T\tr(P_t)$) and the $y$-axis represents how many schedules (out of the $4^{10}$ possible ones) can achieve that cost\footnote{To be precise, we divided the \textit{x}-axis into intervals of 0.5 (e.g., [$90,90.5$] etc.), and evaluated how many schedules produce a cost within each interval.}. 
Such a histogram represents how likely it is to find a random schedule that will produce a given cost.
For this example, our algorithm finds the optimal schedule, whereas the other algorithms fail to find the optimal solution.

\section{Conclusion} \label{S:conclusion}
In this paper, we reformulated the sensor scheduling problem as a sensor design problem whose convex relaxation is solved by a semidefinite programming approach.
While such a relaxation does not readily produce a solution to the scheduling problem, we presented a \textit{covariance-tracking} algorithm to construct a sensor schedule from the solution of the sensor design problem.
The foundation of our algorithm is justified by using an approximate dynamic programming based argument where we show that the tracking based algorithm indeed minimizes an upper bound of the optimal cost (value function).A sub-optimality bound of our proposed algorithm is also derived and discussed.
Performance of our algorithm is demonstrated on several examples and compared with several existing methods to show the merit of our framework.



\bibliographystyle{agsm}
\bibliography{arXiv}             

\end{document}